\documentclass[prd, preprint]{revtex4}
\usepackage{epsfig,graphicx,latexsym}
\usepackage{aas_macros}

\newcommand{\be}{\begin{equation}}
\newcommand{\ee}{\end{equation}}
\newcommand{\bel}[1]{\begin{equation}\label{#1}}
\newcommand{\ba}{\begin{eqnarray}}
\newcommand{\ea}{\end{eqnarray}}
\newcommand{\bal}[1]{\begin{eqnarray}\label{#1}}

\begin{document}

\title[Combining GW Parameter Estimation for Multiple Sources]{Parameter estimation on gravitational waves from multiple coalescing binaries}



\author{Ilya Mandel}
\affiliation{Northwestern University, Evanston, IL  60208, USA}
\email{ilyamandel@chgk.info}
\date{December 30, 2009}

\begin{abstract}
Future ground-based and space-borne interferometric gravitational-wave detectors may capture between tens and thousands of binary coalescence events per year.  There is a significant and growing body of work on the estimation of astrophysically relevant parameters, such as masses and spins, from the gravitational-wave signature of a single event.  This paper introduces a robust Bayesian framework for combining the parameter estimates for multiple events into a parameter distribution of the underlying event population.  The framework can be readily deployed as a rapid post-processing tool.
\end{abstract}

\maketitle

\section{Introduction}

Advanced LIGO and Advanced Virgo, the next generations of the current ground-based gravitational-wave detectors, are scheduled to begin operation around 2015 \cite{LIGO,   Fritschel:2003, AdvLIGOdesign,   Acernese:2008}.  According to the currently available astrophysical predictions, they may detect tens or even hundreds of gravitational waves (GWs) from the coalescence of compact-object binaries during their operation (see, for example, \cite{Kalogera:2004tn, OShaughnessy:2008}).  Meanwhile, the proposed space-borne detector LISA \cite{LISA} could detect GWs from several coalescences of massive black-hole binaries~\cite{Sesana:2004} and hundreds of extreme-mass-ratio inspirals of stellar-mass compact objects into massive black holes~\cite{Gair:2004}. 

Although the first GW detections will be interesting in their own right, ultimately, the search for GWs is carried out in order to glean new astrophysical information about the universe.  Much of this information will come from the statistical distribution of the source parameters, such as masses and spins.  However, the gravitational-wave signal is a complex function of many parameters (up to 17 for a spinning black-hole binary in an eccentric orbit), which is buried in detector noise.  The correlations between the signal parameters and the presence of the statistical noise mean that a single set of best-fit parameters is insufficient to describe a detected GW event; one must also provide estimates of the errors in these parameters in a multi-dimensional space.  Fortunately, Bayesian methods produce not only the most likely set of signal parameters, but also the full posterior probability distribution function (PDF) in the multi-dimensional parameter space.  Techniques such as Markov-chain Monte Carlo (MCMC) and Nested Sampling have already been applied to LIGO/Virgo data (e.g.~\cite{vanderSluys:2008b, vanderSluys:2009, VeitchVecchio:2008}), and to mock LISA data (e.g.~\cite{Wickham:2006, Brown:2007, LittenbergCornish:2009}).  However, previous work has been focused on determining the parameters of a single event, or on simultaneously estimating the parameters of several concurrent signals in the case of LISA \cite{CornishLittenberg:2007, Feroz:2009}.


When gravitational-wave signals from multiple events have been detected, it is highly desirable to combine the parameter estimates from individual detections into a statement about the probability distribution of the relevant parameters in the underlying source population.  This distribution can then be compared, for example, with the predictions from astrophysical models of the population and used to constrain poorly known astrophysical quantities, such as stellar-wind strength or common-envelope efficiency \cite{MandelOShaughnessy:2010}.   Converting individual parameter estimates into a distribution is a necessary step both when estimating parameters for non-overlapping events (LIGO) and for joint parameter estimation on concurrent signals (LISA).  

\begin{figure*}
\includegraphics[keepaspectratio=true, width=0.4\textwidth]{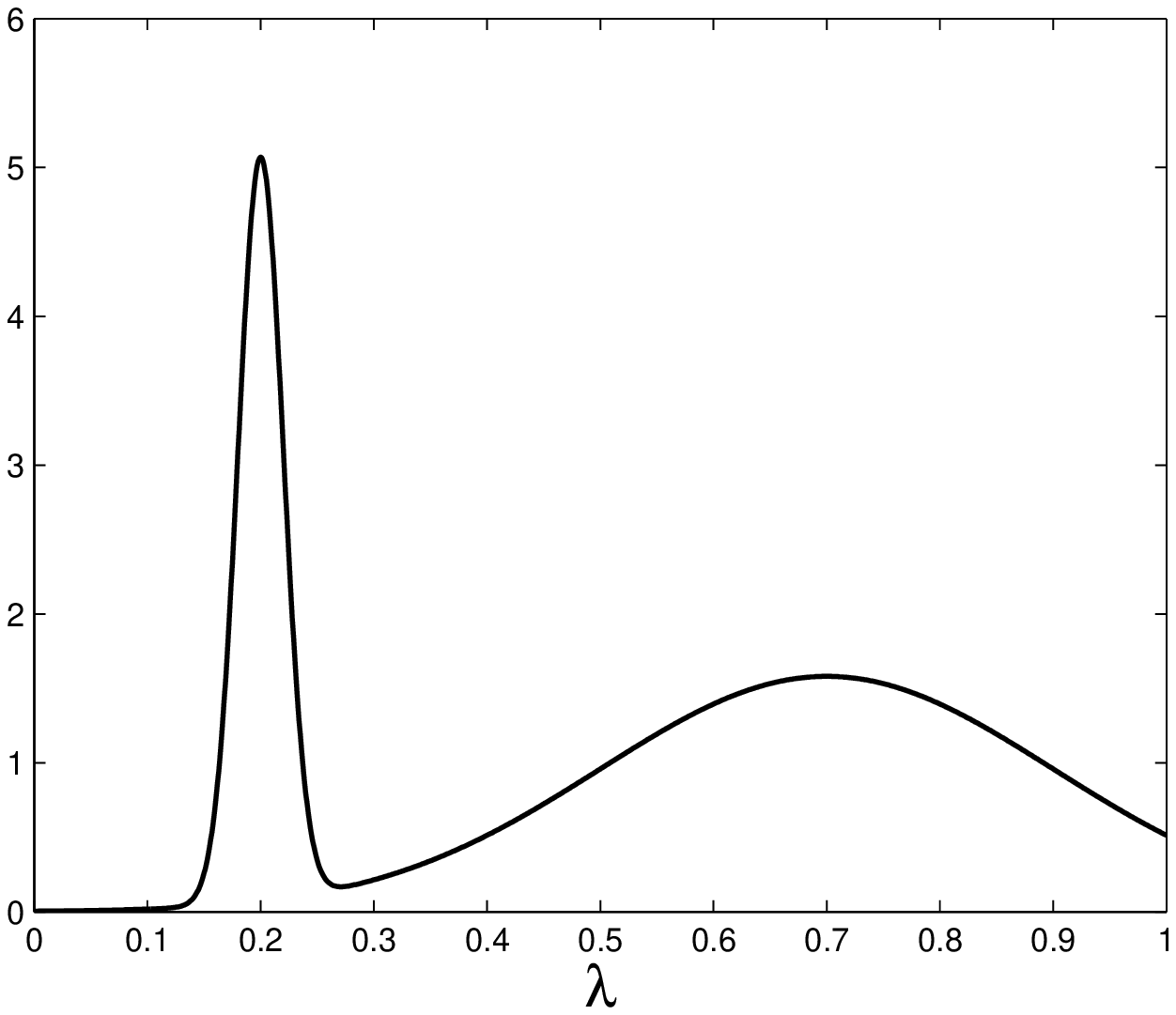}
\hskip0.75in
\includegraphics[keepaspectratio=true, width=0.4\textwidth]{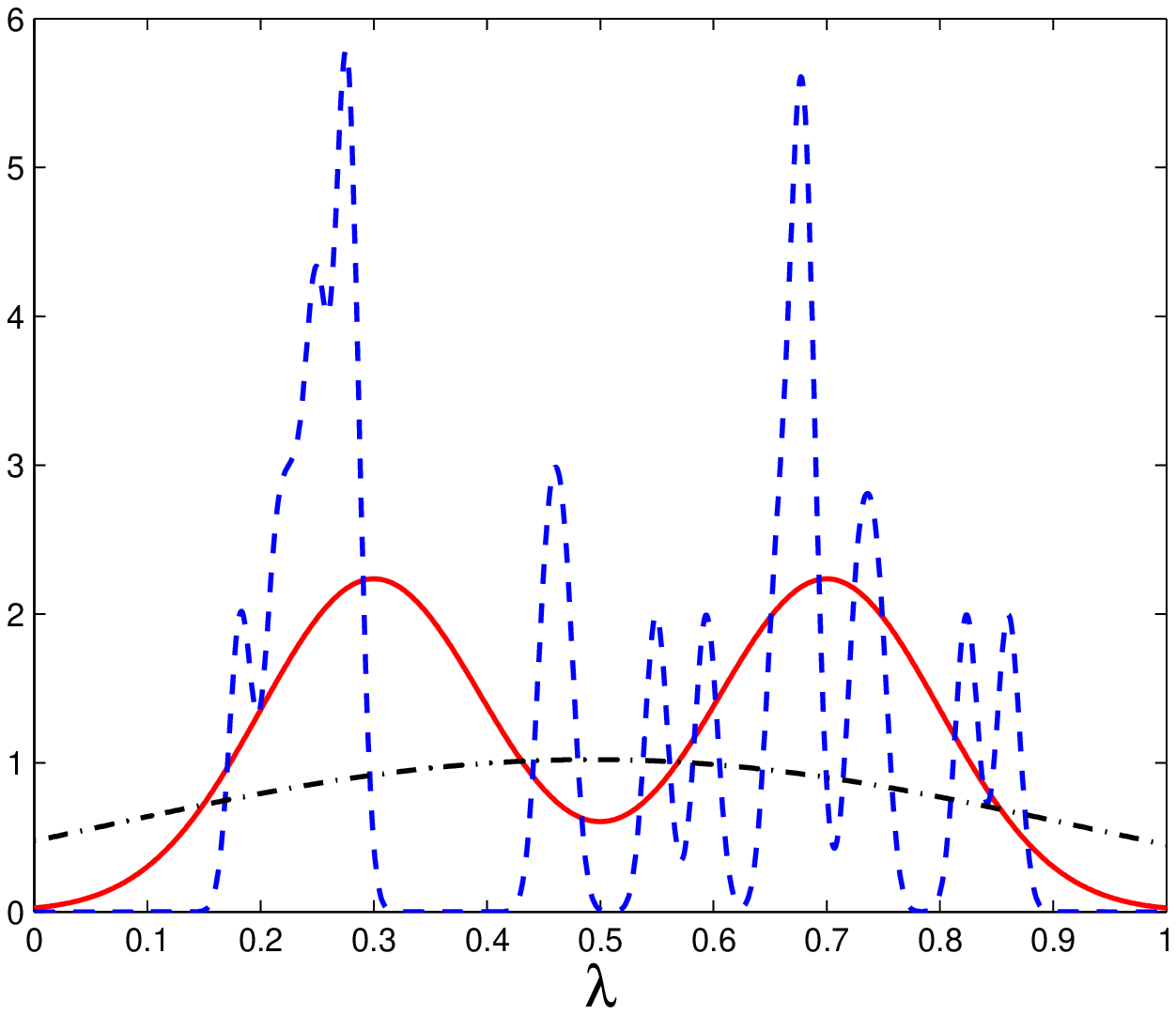}
\vskip-0.1in
\caption{
The left panel shows a possible multimodal marginalized PDF for a single parameter of a single detected event.  The right panel shows the results of estimating the distribution from individual samples by means of a kernel density estimator.  Twenty samples are taken from a bimodal distribution (solid red); the blue dashed curve shows a KDE where the smoothing parameter (bandwidth) was probably chosen to be too small, while the black dash-dotted curve shows the effect of a KDE with a large bandwidth parameter that over-smoothes the distribution.
  }
  \label{fig:KDE}  
\end{figure*}

It is possible to make non-parametric fits to the observations, resulting in a smooth estimate for the probability density function of the distribution, by means of a kernel density estimator (KDE) \cite{Parzen:1962}.  However, the parameter estimates of individual events could be multimodal due to partial waveform degeneracies in the multi-dimensional parameter space and the effects of instrumental noise, as indicated in the left panel of Fig.~\ref{fig:KDE}, and it would be necessary to address the problem of properly folding non-trivial single-event PDFs into the KDE.  There is also a more serious problem with the strong dependence of the KDE on the choices of kernel and bandwidth, as indicated in the right panel of Fig.~\ref{fig:KDE}.  There is a danger of either over-emphasizing the individual samples or over-smoothing the modes of the distribution, without an obvious mechanism for selecting the free parameters.  Therefore, we choose to retain full control over the fitting process in a way that does not require any ad hoc choices for smoothing parameters.

We accomplish this by developing a fully Bayesian framework for extracting a population distribution from a set of individual detections.  We assume that the parameters of the individual events have been successfully estimated and posterior PDFs have been computed, and use Bayesian techniques to derive statements about the distribution of parameters in the sampled population.  We derive the theoretical framework for this calculation in Section II, address some important practical issues related to the calculation in Section III, describe a specific implementation in Section IV and add some closing comments in Section V. 

\section{Theoretical framework}

Consider a data stream $d(t)$, which consists of a gravitational wave $h(t)$ from a source with true parameters $\vec{\theta}_{T}$ and stationary, Gaussian noise $n(t)$: $d(t)=h(\vec{\theta}_{T}; t)+n(t)$.  The parameters $\vec{\theta}$ encode the masses of the inspiraling binary components, their spin magnitudes and directions, the location of the binary on the sky, the luminosity distance to the binary, the orientation of the orbit, the time of coalescence and several phase variables.  The likelihood for this data given a model $h(\vec{\theta})$ is
\bel{likelihood}
{\cal L}(d|\vec{\theta}) \propto e^{-\frac{\left\langle d - h(\vec{\theta}) | d - h(\vec{\theta}) \right\rangle}{2}}
	\propto \exp \left\{ \left\langle d|h(\vec{\theta})\right\rangle 
		- \frac{1}{2} \left\langle h(\vec{\theta})|h(\vec{\theta})\right\rangle\right\}.
\ee
Here, 
\bel{overlap}
\langle a|b \rangle = 4 \int_0^\infty \frac{\tilde{a}(f) \tilde{b}^*(f)}{|S_n(f)|}df
\ee
is the overlap of the Fourier transforms of the signals and $S_n(f)$ is the one-sided power spectral density of the noise.  In a network of multiple detectors with uncorrelated noise (or multiple uncorrelated data streams in the case of LISA), the overall likelihood is given by the product of the individual detector likelihoods; we work with a single detector for simplicity of notation. 

The posterior PDF is given by the Bayes theorem:
\bel{posterior}
p(\vec{\theta}|d) = \frac{p(\vec{\theta}) {\cal L}(d | \vec{\theta})}{p(d)},
\ee
where $p(\vec{\theta})$ is the prior on the parameters and $p(d)$ is the evidence (which can be ignored, like the factor of $\exp(\langle d|d\rangle)$ in Eq.~(\ref{likelihood}), when considering the relative posteriors for two parameter sets).  The prior could be set based on astrophysical expectations (e.g., flat priors on the sky location under the assumption that sources are distributed isotropically).  
Posterior PDFs can be marginalized over all but one parameter to produce single-parameter PDFs.  For example, Figure 1 of \cite{vanderSluys:2009} shows an example of posterior PDFs of a LIGO hardware injection as computed with the SPINspiral MCMC-based parameter-estimation code.

We assume that $k$ events have been detected and their associated parameter estimates are independent.  This may not be strictly true for LISA events, since multiple events overlap in the data stream, so the parameter estimation is correlated. On the other hand, the LIGO/Virgo events last only a few seconds, and so should indeed be uncorrelated, although any systematic biases (e.g., those arising from imperfect waveform modeling) could create correlations.  We will focus on the distribution of a single parameter $\lambda$, assuming all other parameters have been marginalized over, although our approach extends trivially to joint distributions of multiple parameters.  The question we want to answer is: what population distribution $f(\lambda)$ is most consistent with the observed samples? In other words, we want to know the conditional probability $p(f(\lambda)|p_1(\lambda),...p_k(\lambda))$, where $p_i(\lambda)$ is the posterior PDF for $\lambda$ from measurement $i \in [1,k]$, marginalized over all other parameters.  

Note that we are obtaining a {\it distribution of distributions}: we can not uniquely identify the population distribution from $k$ samples, but rather, we can describe the relative probability that the measurements are consistent with a given choice of the population distribution.  We apply Bayes theorem to this conditional probability in the usual way:
\bel{posteriordist}
p\left(f(\lambda)|p_1(\lambda),...p_k(\lambda)\right) = \frac
	{p\left(p_1(\lambda),...p_k(\lambda)|f(\lambda)\right)\ \pi(f(\lambda))}
	{p\left(p_1(\lambda),...p_k(\lambda)\right)}.
\ee
Here, the prior $\pi(f(\lambda))$ should include any a priori astrophysical understanding.  

The key ingredient of Eq.~(\ref{posteriordist}) is the likelihood of obtaining the observed set of individual PDFs given the underlying population distribution $f(\lambda)$.  Because all $k$ draws are assumed to be independent, the probability of obtaining a particular set of observations is equal to the product of the individual probabilities for obtaining a particular observation \footnote{After developing this approach, we learned that a similar technique has been applied to the distribution of exoplanetary orbits in \cite{FabryckyWinn:2009}.}:
\be
p\left(p_1(\lambda),...p_k(\lambda)|f(\lambda)\right) = \prod_{i=1}^k p\left(p_i(\lambda)|f(\lambda)\right).
\ee
For a discrete set of parameter values and a precise observation, the probability of drawing a particular value is just given by the probability density of that value in the population distribution.  If a box contains numbered balls, the probability of drawing ball number $n$ is equal to the fraction of balls in the box bearing that number.  On the other hand, if one imagines that after the ball is drawn, there are several possibilities for the observation, then  the probability of drawing that ball is the sum over all relevant $n$ of the product of the probability density and the observed probability of the draw.  E.g., if there's a 50\% chance that the chosen ball is a $6$ and 50\% that it's a $9$ because of ambiguous labeling, the likelihood of this draw is $0.5 \times f(6)+0.5 \times f(9)$.  In the case of a continuous distribution of parameter values, the sum is replaced by an integral, so that the probability of obtaining a particular observed posterior PDF given the population distribution $f(\lambda)$ is:
\bel{PDFprobability}
p\left( p_i(\lambda)|f(\lambda) \right) = \int_{{\rm all}\ \lambda}  p_i(\lambda) f(\lambda) d\lambda.
\ee

We must be very careful to include selection effects in this analysis.  For example, the sensitivity of the LIGO-Virgo network is a function of the chirp mass of the coalescing binary \cite{BradyFairhurst:2008}.  In general, we will be fitting to a distribution of {\it detectable} events, not to the distribution of the underlying source population, unless these selection effects are properly folded into $p(f(\lambda))$.  The precise nature of the selection effects will depend on the details of the employed detection pipeline.   Appropriately quantifying the selection bias will therefore require the analysis of a large set of injections into realistic detector noise with the full detection pipeline, and is outside the scope of this paper.

The space of possible functional forms of the population probability distributions $f(\lambda)$ is infinite, so we must select a suitable parametrization for $f(\lambda)$.  One simple option is a histogram parametrization: the space of possible values of $\lambda$ can be divided into some small number $n$ of bins, so that assuming a flat probability distribution within each bin, $f(\lambda)$ can be be described by $n-1$ parameters listing the probability within each bin (the probability of the $n$th bin is set by the overall normalization requirement).  Alternatively, a multimodal Gaussian distribution can be used, parametrized by the mean and standard deviation of each Gaussian.  Of course, the specific choice of the parametrization should depend on astrophysical understanding of the given parameter.  For instance, for compact-object masses of LIGO-Virgo coalescences, two Gaussians may be suitable, one each for neutron-star and black-hole masses -- or perhaps three, if the black-hole masses have two different distributions depending on whether the system is formed dynamically in a dense stellar environment or via isolated binary evolution in the field \cite{MandelOShaughnessy:2010}.

Clearly, more parameters in $f(\lambda)$ will tend to increase the likelihood in Eq.~(\ref{posteriordist}).  However, Occam's razor dictates that the fewest possible number of parameters should be used.  Bayesian model selection provides a quantitative method of determining the optimal number of model parameters, such as the number of bins in a histogram parametrization or the number of Gaussians in a multimodal parametrization.  The odds ratio of two models $M_1$ and $M_2$ is equal to
\bel{odds}
{\cal O} = \frac{p(M_1)}{p(M_2)} \frac
{\int p\left(p_1(\lambda),...p_k(\lambda)|f(\lambda); M_1 \right)\ \pi(f(\lambda)| M_1)\ d\lambda} 
{\int p\left(p_1(\lambda),...p_k(\lambda)|f(\lambda); M_2 \right)\ \pi(f(\lambda)| M_2)\ d\lambda} ,
\ee
where the first factor is the prior ratio for the two models, which can be set to $1$ in the absence of other information, and the second factor, known as the Bayes factor, is the ratio of marginal likelihoods or evidences.  The Bayes factor can thus be computed by integrating Eq.~(\ref{posteriordist}) over all values of $\lambda$ for each competing model.  Strictly speaking, consistency requires that the same probability distribution $f(\lambda)$ be used in setting the priors $p(\vec{\theta})$ in Eq.~(\ref{posterior}) for computing the individual-event posterior PDFs $p_i(\lambda)$.  This is likely to be computationally impractical for model selection.  However, our experience suggests that the posteriors on individual parameters do not depend strongly on the priors when the data has sufficient discriminating power.  We can test this hypothesis by computing the distribution of $f(\lambda)$ using pre-computed $p_i(\lambda)$, and then re-evaluate these single-parameter PDFs using the new priors to see if there is a significant effect.

\section{Practical considerations}

The approach outlined above should make it possible to determine the probability distribution function $p(f(\lambda))$ for the parameters of an assumed population distribution $f(\lambda)$, given a set of single-event posterior PDFs $p_i(\lambda)$.  However, while the distribution $f(\lambda)$ is a parametrized analytical distribution, the single-event PDFs will be sampled numerically (for instance, by a Markov Chain Monte Carlo technique) and will be discrete.  This presents a technical challenge when evaluating the  integral in Eq.~(\ref{PDFprobability}).

If a histogram ansatz is taken for the distribution $f(\lambda)$, this is not an issue, since the value of the discretely sampled PDF $p_i(\lambda)$ in a given bin of the histogram is just the fraction of all points of the PDF sample that fall within that bin.  Alternatively, if each of the observations $p_i(\lambda)$ falls within a single bin, our belief about the population distribution $p(f(\lambda))$ can be read off directly as the Dirichlet distribution (this is generally the case whenever the parameter $\lambda$ can be suitably discretized).

Another possibility is to make an analytical fit to the discrete PDF, and then use the results of the fit in evaluating Eq.~(\ref{PDFprobability}).  One simple possibility, which should be appropriate for most sharply-peaked, single-mode PDFs, is a Gaussian fit.  However, any errors in this fit due to using an incorrect analytical form for the fit can lead to biases in the resulting $p(f(\lambda))$ that will be difficult to quantify.  

Instead, we choose the following approach.  We note that, by construction, the density of points in a PDF that is discretely sampled according to the posterior is proportional to the value of the PDF.  Therefore, the typical separation $d\lambda$ between neighboring points in the PDF scales as $1/p(\lambda)$.   If we rescale the parameter $\lambda$ so that its allowed prior range is $[0,1]$, and the PDF consists of $N$ discrete points, we can approximate $d\lambda$ as
\be
d\lambda = \frac{1}{N} \frac{\int p(\lambda) d\lambda}{p(\lambda)}.
\ee
Unlike in the computation of the Bayes factor, which requires the marginal likelihood (evidence) to be calculated, here we are only interested in the distribution of points of the sampled PDF, not the actual values of the posterior $p(\lambda)$; we can therefore ignore the factor $\int p(\lambda) d\lambda$ which does not influence the relative values of likelihoods for different distributions $f(\lambda)$.
This allows us to approximate the integral in Eq.~(\ref{PDFprobability}) as
\bel{PDFprob-discrete}
p\left( p_i(\lambda)|f(\lambda) \right) = \frac{1}{N_i} \sum_{j=1}^{N_i} f(\lambda_i^{(j)}),
\ee
where $\lambda_i^{(j)}$ is the $j$th element of the discrete chain representing the posterior PDF of the $i$th observed event, $p_i(\lambda)$.  Thus, the posterior for a distribution $f(\lambda)$ given $k$ single-event PDFs is
\bel{distfinal}
p\left(f(\lambda)|p_1(\lambda),...p_k(\lambda)\right) \propto \pi(f(\lambda))\ 
	\prod_{i=1}^k \left[ \frac{1}{N_i} \sum_{j=1}^{N_i} f(\lambda_i^{(j)}) \right],
\ee
where we ignored the evidence since we are only interested in the relative posteriors.  Potentially, this approach could suffer from errors in regions of low PDF $p(\lambda)$ which are poorly sampled, although the contribution of these regions to the integral should not be significant.  We can estimate the accuracy of the computation via bootstrapping techniques, as described in Section \ref{sec:example}.
 
Evaluating single-parameter or multi-parameter distributions as described here will require us to store a database of the full multi-dimensional PDFs sampled according to the posterior, which can then be queried for specific follow-ups.  This is already being done by Bayesian follow-up codes in the LIGO-Virgo compact-binary-coalescence pipeline.  Of course, the faithfulness of the single-event PDFs are a concern.  In addition to possible errors from imperfect sampling of potentially multi-modal distributions -- always an issue with Markov Chain Monte Carlo -- systematic errors are possible due to imperfect waveform knowledge and the use of approximate waveform families for parameter estimation \cite{CutlerVallisneri:2007, Huerta:2009}.  The importance of such systematic errors is particularly great in cases when the statistical errors are expected to be small due to high signal-to-noise ratios, such as for supermassive black-hole binary coalescences observable with LISA.






Another interesting question is the number of individual measurements necessary to make interesting astrophysical statements based on the parameter distributions.  Bulik and Belczynski estimated that as few as $O(20)$ chirp-mass measurements could distinguish between some alternative population-synthesis models, though the required number could grow to $O(1000)$ for models yielding more similar predictions.
However, this analysis was based on the assumption of perfect measurements, and uncertainties in the individual PDFs will tend to increase the minimum sample size.  On the other hand, folding in other observable parameters besides the chirp mass (e.g., the mass ratio) should aid in resolving between alternate astrophysical models.

\section{Sample calculation \label{sec:example}}

In this section, we demonstrate an example of the calculation of the population probability distribution  from a set of imperfectly estimated samples using the methodology described above. 

For our example, we imagine that the true distribution is a Gaussian in the single parameter $\lambda$.  We assume that the allowed prior range on $\lambda$ is $[0,1]$, and we arbitrarily choose mean $\mu_{\rm true}=0.4$ and standard deviation $\sigma_{\rm true}=0.1$ as the parameters of the true distribution.  We further imagine that $k$ events have been detected, i.e., we choose $k$ samples according to the true distribution.  For each detected event, we assume that the parameter of interest is not known perfectly, but is represented by a posterior PDF.  In this case, we model each PDF as a Gaussian with an unbiased mean centered on the chosen value of the sample, but with a standard deviation uniformly chosen in the range $[a, b]$ to represent the statistical errors in parameter estimation.  In order to produce a discrete PDF for each event, we sample this PDF with $N$ points.   (We cut off all distribution functions with non-zero support outside the interval $\lambda \in [0,1]$ at the edges of the allowed range of $\lambda$.)  For the default case, we pick $k=20$ events from the true distribution.  Each event has a PDF with a standard deviation uniformly chosen in the range $[a,b]=[0, 0.2]$, sampled with $N=100$ points.  

To reconstruct the distribution $f(\lambda)$, we assume that the detected events are sampled from a Gaussian distribution with flat priors on its mean $\mu$ and standard deviation $\sigma$.  We fit for $\mu$ and $\sigma$ with a Markov Chain Monte Carlo, using Eq.~(\ref{distfinal}) to evaluate the relative posteriors for different choices of $\mu$ and $\sigma$.  Since this is a fairly simple two-dimensional problem, we use a straightforward Metropolis-Hastings implementation.  The jump-proposal distribution is just a uniform sampling in the prior.  We find that $100000$ points are more than sufficient to accurately sample the parameter space.  We throw away the first $20\%$ of the chain as the burn-in (an over-cautious choice).  

\begin{figure*}
\includegraphics[keepaspectratio=true, width=0.3\textwidth]{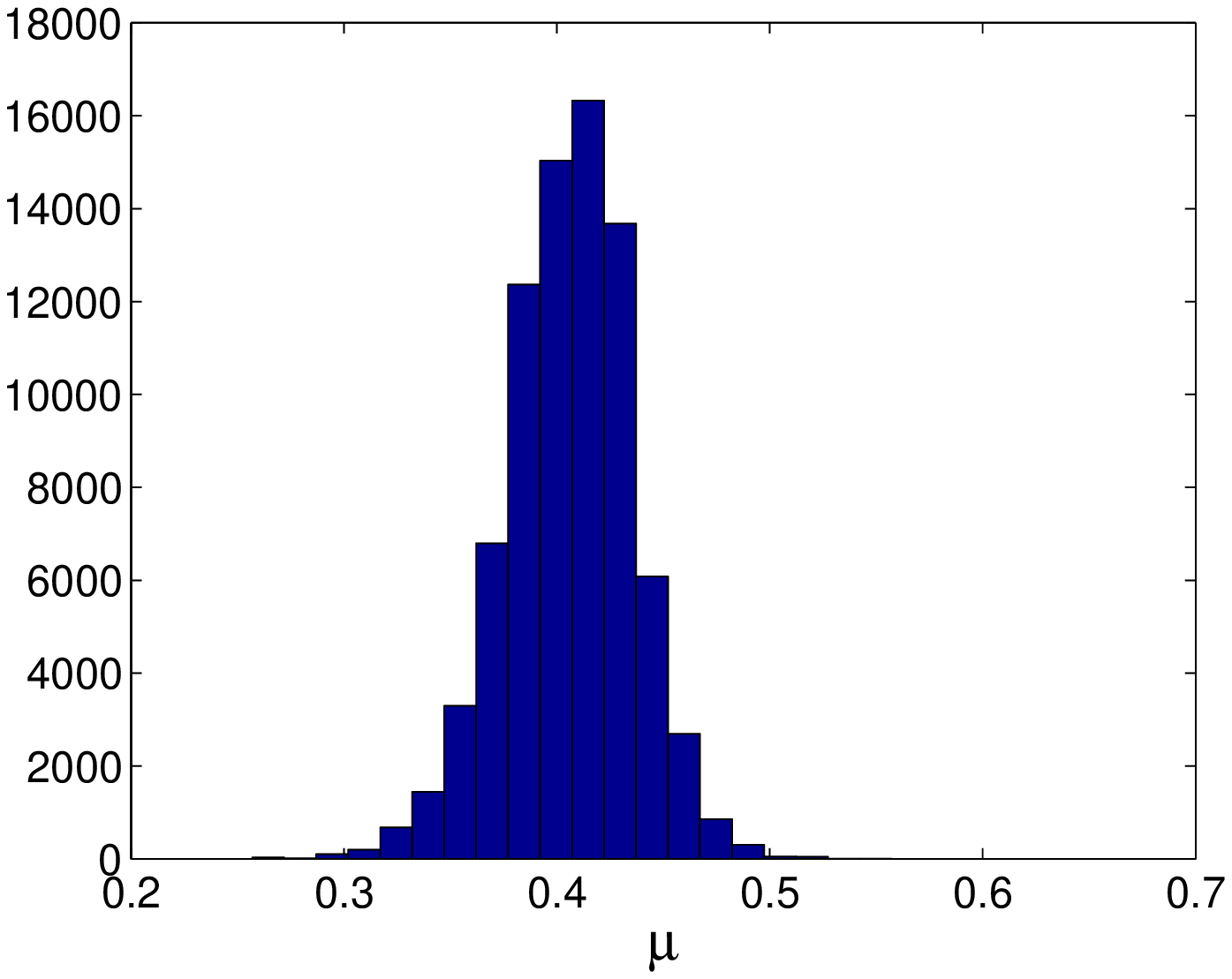}
\hskip0.25in
\includegraphics[keepaspectratio=true, width=0.3\textwidth]{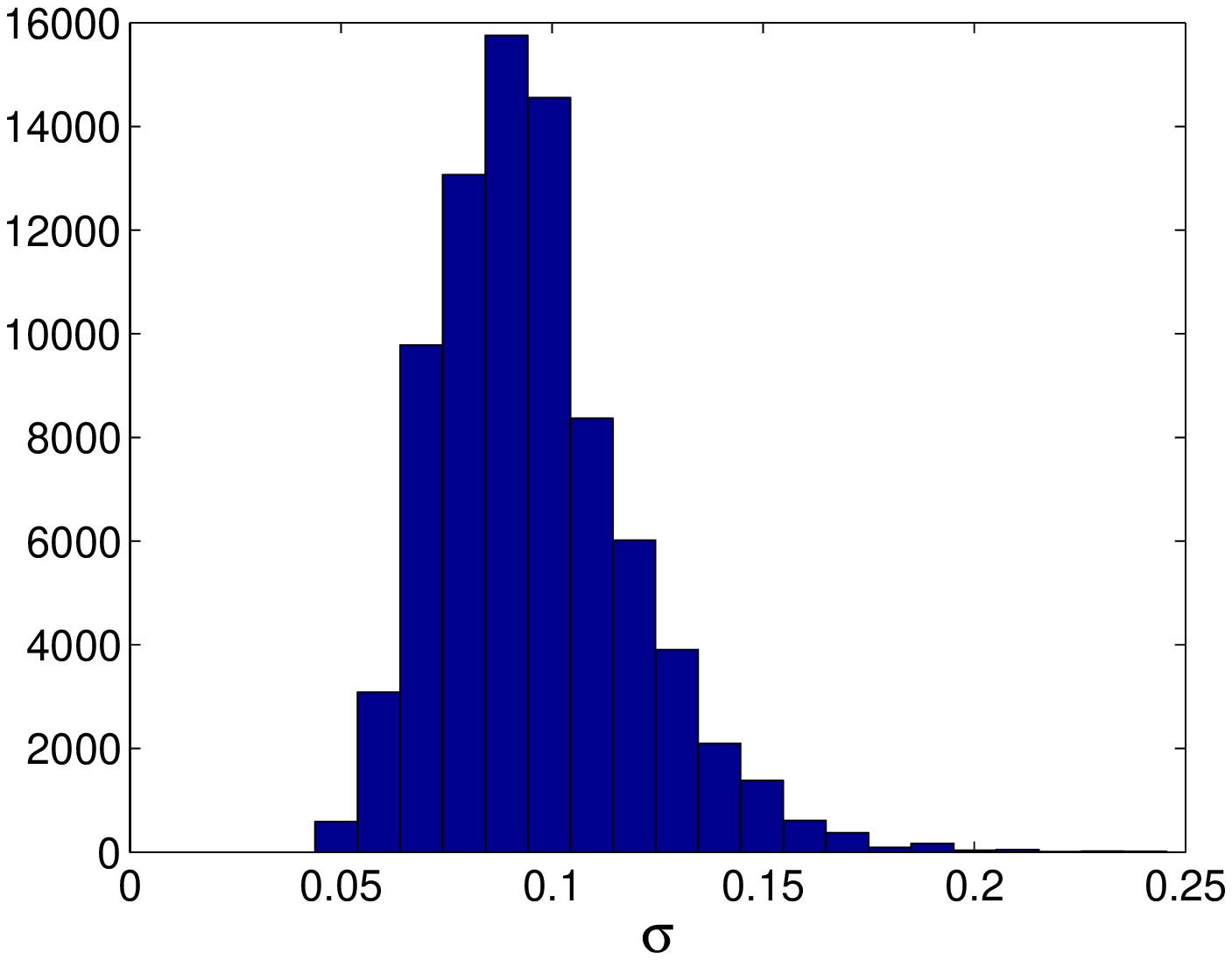}
\hskip0.25in
\includegraphics[keepaspectratio=true, width=0.3\textwidth]{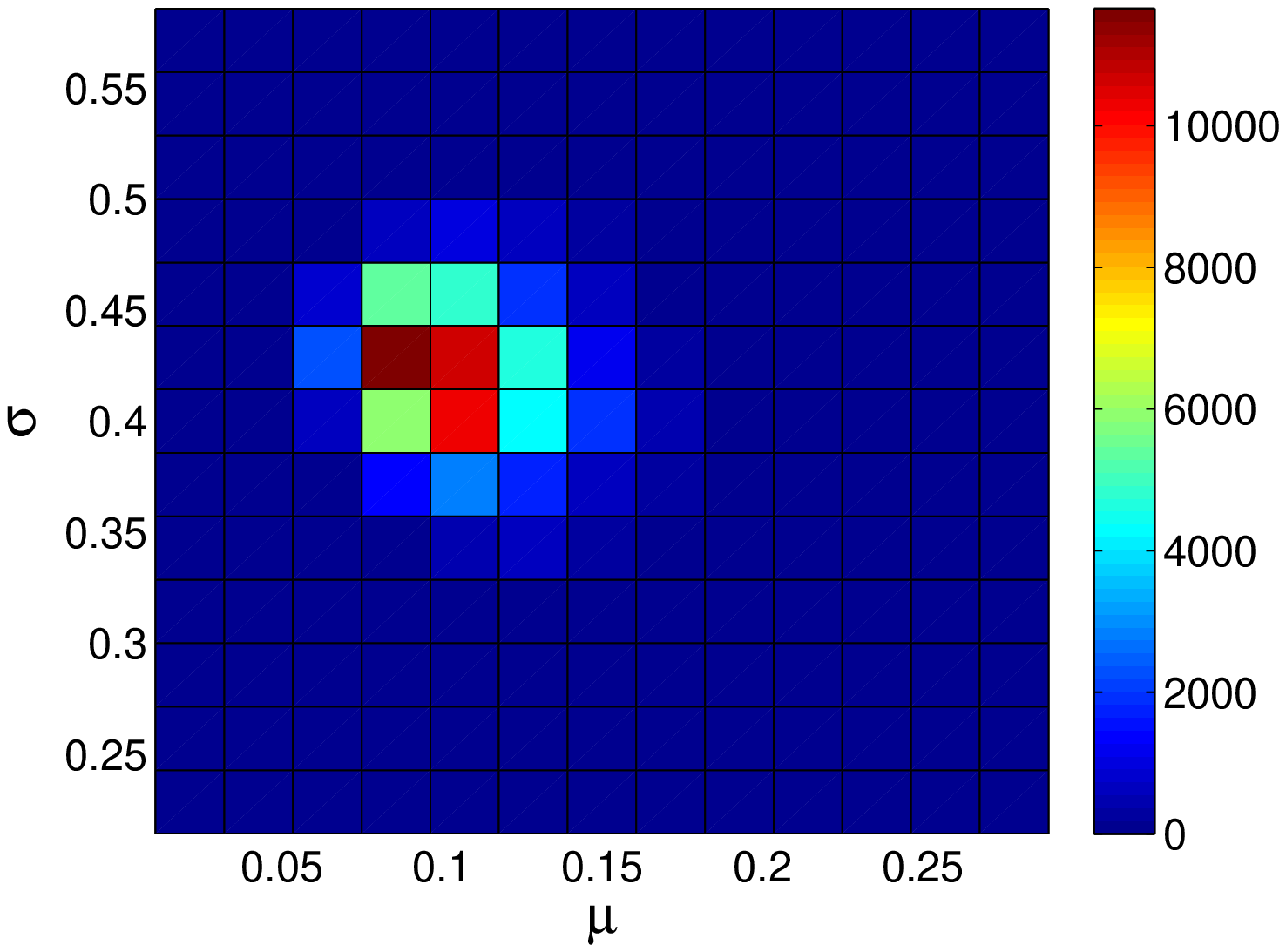}
\vskip-0.1in
  \caption{ Recovered one-dimensional posterior probability distributions for mean $\mu$, standard deviation $\sigma$, and their join two-dimensional distribution.  The true probability distribution with $\mu_{\rm true}=0.4$ and  $\sigma_{\rm true}=0.1$ was sampled with $k=20$ detected events.  The PDF obtained for each event is a Gaussian centered on its true value with standard deviation taken uniformly from $[0, 0.2]$, sampled with $N=100$ points.  The weighted averages of the recovered parameters over the posterior density of the reconstructed population distribution are $\bar{\mu}=0.4055$ and $\bar{\sigma}=0.0958$, respectively.}
\label{fig:default}  
\end{figure*}

Fig.~\ref{fig:default} shows the resulting one-dimensional marginalized posteriors for $\mu$ and $\sigma$ and the joint two-dimensional posterior.  With these choices, the distribution-fitting code takes less than a minute to run on a modern laptop.  We study the accuracy of the reconstruction by separately varying the chosen values of $k$, $N$, and $[a,b]$ and comparing with the default case.  

Increasing the number of samples $N$ of each PDF beyond $100$ does not appear to increase the accuracy with which the distribution is reconstructed, although the accuracy drops for $N\lesssim 100$. Since we expect at least $100$ uncorrelated points in the typical PDF chain (SPINspiral runs typically have $\sim10^3-10^4$ uncorrelated points \cite{vanderSluys:2009}), the number of PDF samples for each event does not appear to be a concern. We used a fairly simple PDF model here, however, so the minimal value of $N$ may need to be greater in exceptional cases of multi-modal PDFs with widely spaced modes.

\begin{figure*}
\includegraphics[keepaspectratio=true, width=0.4\textwidth]{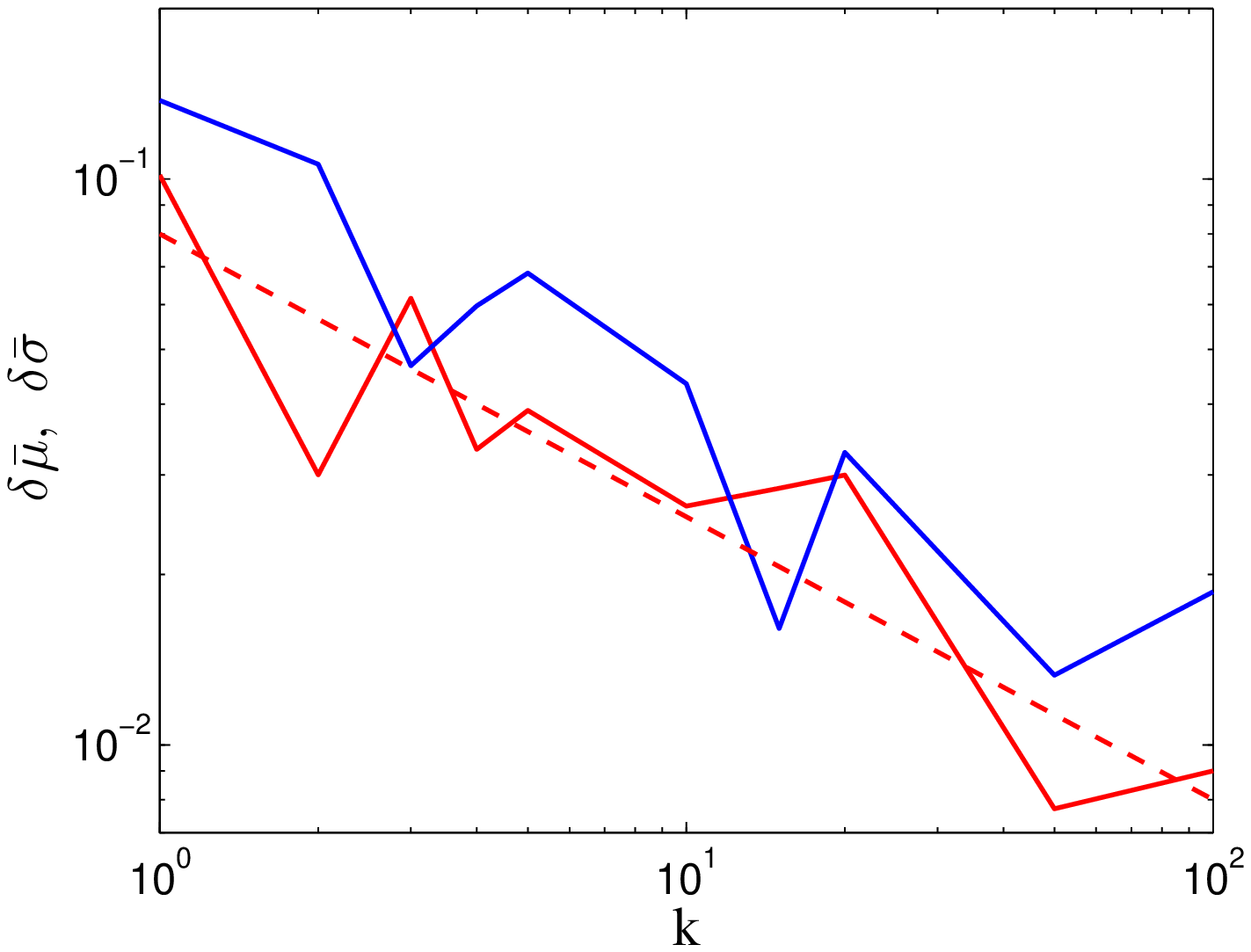}
\hskip0.75in
\includegraphics[keepaspectratio=true, width=0.4\textwidth]{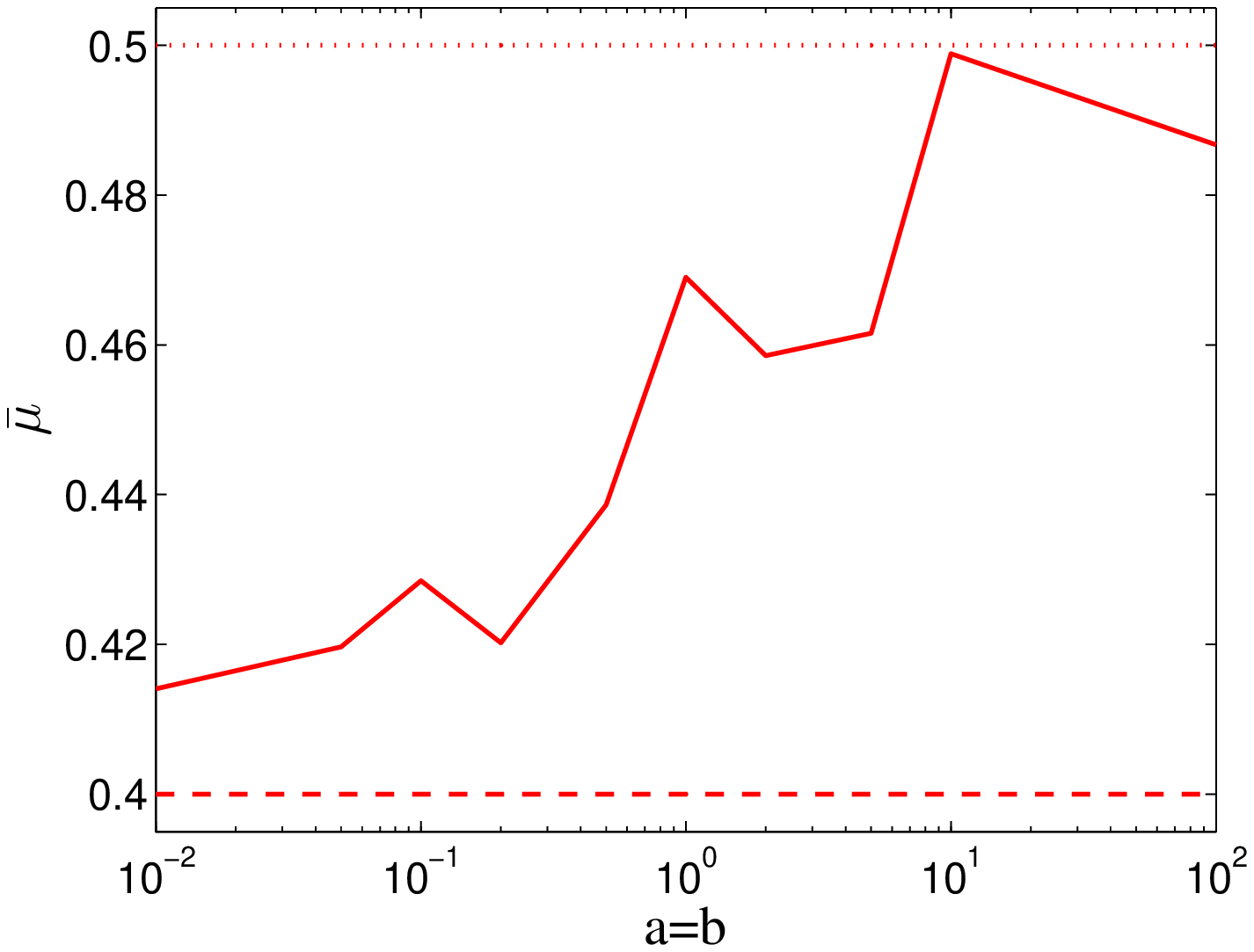}
\vskip-0.1in
  \caption{The left panel shows the error in the recovered mean of the population distribution $\delta \bar{\mu} = \bar{\mu}-\mu_{\rm true}$ (red) and the standard deviation  $\delta \bar{\sigma} = \bar{\sigma}-\sigma_{\rm true}$ (blue) as a function of the number of detected events $k$.  The dashed red curve is the approximate analytic fit $\delta \bar{\mu} \propto 1/\sqrt{k}$.  The right panel shows the estimated mean $\bar{\mu}$ as a function of the statistical uncertainty in each event $a=b$.  The dashed line is the true mean of the population distribution $\mu_{\rm true}=0.4$, while the dotted line is the mean of the prior on $\mu$.}
\label{fig:kab}  
\end{figure*}

We next consider the effect of the number of detected events $k$ on the quality of the reconstruction of the population distribution function.  As in the default run, we model the statistical errors by discretizing each single-event PDF with $N=100$ points drawn from a distribution with a standard deviation uniformly chosen from $[a,b]=[0,0.2]$.  We consider $10$ different choices for the number of detected events $k=1, 2, 3, 4, 5, 10, 15, 20, 50, 100$ and compute the posterior probability distribution for the parameters mean $\mu$ and standard deviation $\sigma$ for each value of $k$.  The left panel of Fig.~\ref{fig:kab} shows the weighted averages of these parameters $\bar{\mu}$ and $\bar{\sigma}$ over the posterior.  Because the errors in the mean and standard deviation of a given run are random variables, affected by the particular sample of detected events, we average the errors in the estimates of the mean $\delta \bar{\mu} = \bar{\mu}-\mu_{\rm true}$ and standard deviation  $\delta \bar{\sigma} = \bar{\sigma}-\sigma_{\rm true}$ over $5$ runs for each data point in the figure.

There is a clear anti-correlation between the number of detected events and the errors in the recovered parameters of the population distribution: as expected, the fewer detections there are, the harder it is to accurately extract the distribution they sample.  In fact, the correlation between 
$\log k$ and $\log(\delta \bar{\mu})$ is $-0.89$, and the correlation between $\log k$ and $\log(\delta \bar{\sigma})$ is $-0.90$.  If the parameters of every detected event were measured perfectly, i.e., if single-event PDFs were delta-fucntions, then the error in the recovered mean should scale inversely with the square root of the number of detected events.  The data for $\delta \bar{\mu}$ in Fig.~\ref{fig:kab} is consistent with this trend.  We also note that for a single detected event $k=1$, the typical error is consistent with the standard deviation of the true distribution $\sigma_{\rm true}$, as expected.  Meanwhile, for low values of $k$, the standard deviation of the population distribution is significantly over-estimated because we are sensitive to the variance in the individual PDFs, which can be greater than the variance in the population distribution for the chosen parameters.

Finally, we consider the impact of the statistical uncertainty in the evaluation of single-event PDFs.  For this comparison, we keep the number of detected events fixed at $k=20$ and the number of points per single-event PDF fixed at $N=100$.  In order to model the statistical uncertainty in each measurement, we vary the standard deviations of the single-event PDFs, assigning to all PDFs in a given run the same standard deviation $a=b$.  In the right-hand panel of Fig.~\ref{fig:kab}, we present the results for $10$ different choices of this statistical uncertainty parameter: $a=b=0.01, 0.05, 0.1, 0.2, 0.5, 1.0, 2.0, 5.0, 10.0, 100.0$.    As expected, the error in the estimated mean $\delta\bar{\mu}$ is small and dominated by the randomness of the chosen sample of detected events as long as the statistical uncertainty for each detected event is less than the standard deviation of the population distribution, $a=b \lesssim \sigma_{\rm true} = 0.1$.  The error grows rapidly at that point, and the estimated mean $\bar{\mu}$ asymptotes to $0.5$ for very large values of $a=b$, as the flat PDFs yield no information, so that the posterior of the distribution parameters is set by the prior, whose mean is $0.5$ while $\mu_{\rm true}=0.4$.


\section{Concluding remarks}

In this paper, we have introduced a robust Bayesian technique for estimating a modeled population distribution based on an arbitrary number of independent detected events, each with an associated posterior probability distribution function describing the degree of statistical uncertainty in parameter estimation.  This technique can be readily deployed as an inexpensive post-processing tool, which can run in a tiny fraction of the time required to sample the PDFs of individual events.  

We have demonstrated the application of this technique on a fairly simplistic model of the population distribution and the PDFs of detected events.  However, it is important to point out that because the algorithm does not make any assumptions about the shapes of the single-event PDFs or attempt any analytical fits to the discrete PDFs, the evaluation should be just as easy for arbitrarily complicated (e.g., multimodal) PDFs.  Moreover, Markov Chain Monte Carlo techniques should allow us to estimate the parameters of more complex population-distribution models without additional complications.

On the other hand, a number of issues will require further investigation.  We commented on the importance of appropriately correcting for any selection biases in the detected sample in order to reconstruct an unbiased population distribution, but the actual corrections, which depend on the implementation of the detection pipeline, were outside the scope of this paper.  Similarly, additional work is necessary to characterize any systematic errors in parameter estimation for individual events.  
We have suggested how model selection should proceed in principle, but the choices of particular families of competing models must depend on prior astrophysical knowledge.  Astrophysical inputs are also needed to characterize the number and accuracy of detections which would be necessary to constrain astrophysical models with the observed population distributions of specific parameters such as masses and spins.

\section*{ACKNOWLEDGMENTS}

I am grateful to P.~Brady, J.~Burguet-Castell, W.~Del Pozzo, W.~Farr, J.~Gair, V. Kalogera, R.~O'Shaughnessy, V.~Raymond, M.~van der Sluys, R.~Vaulin, A.~Vecchio, and J.~Veitch for useful discussions.  I acknowledge support from NSF grant PHY-0653321 to Northwestern University and the NSF Astronomy and Astrophysics Postdoctoral Fellowship under award AST-0901985. 

\bibliography{../bibliography/Mandel}

\end{document}